# DV-DVFS: Merging Data Variety and DVFS Technique to Manage the Energy Consumption of Big Data Processing

HOSSEIN AHMADVAND, Sharif University of Technology, Tehran, Iran
FOUZHAN FOROUTAN, Sharif University of Technology, Tehran, Iran
MAHMOOD FATHY, Iran University of Science and Technology and School of Computer Science, Institute for Research in Fundamental Sciences, Tehran, Iran

The email address of the corresponding author: ahmadvand@ce.sharif.edu

**Abstract.** Data variety is one of the most important features of Big Data. Data variety is the result of aggregating data from multiple sources and uneven distribution of data. This feature of Big Data causes high variation in the consumption of processing resources such as CPU consumption. This issue has been overlooked in previous works. To overcome the mentioned problem, in the present work, we used Dynamic Voltage and Frequency Scaling (DVFS) to reduce the energy consumption of computation. To this goal, we consider two types of deadlines as our constraint. Before applying the DVFS technique to computer nodes, we estimate the processing time and the frequency needed to meet the deadline. In the evaluation phase, we have used a set of data sets and applications. The experimental results show that our proposed approach surpasses the other scenarios in processing real datasets. Based on the experimental results in this paper, DV-DVFS can achieve up to 15% improvement in energy consumption.

1. INTRODUCTION

In recent years, due to the huge amount of generated data and the high volume of processing capacity, the power, and energy management of this huge processing, is a considerable value. The authors in [1] estimate that over 40% of the budget in a data center has been spent on electrical power and cooling in 2008. Paying attention to the data variety can prevent the loss of resources such as energy [2]. Also, several approaches addressed this issue, but, in these works, the impact of data variety on the waste of energy has been ignored. Some recent works considered the energy consumption in MapReduce-like distributed processing frameworks [3], [4]. The goal of this researches was to determine the number of worker nodes used by a Hadoop cluster to minimize energy consumption and at the same time guarantee a specified deadline. However, they assumed that they know the input data beforehand. This assumption is not valid in a real environment, so the applicability of such techniques is limited. Furthermore, multiple schedulers consider the deadline constraints [5]; nevertheless, none of the schedulers supports the minimization of energy consumption. The authors in [6] have presented a framework for Efficient Energy Scheduling of Spark workloads. The authors in [7] have used load balancing to improve energy efficiency. They have used the heuristic method for their goal.
Data variety causes some variation in generating results and resource utilization. None of the previous works paid attention to this issue. So, a novel approach can be presented in using data variety for reducing energy consumption.

**Motivation.** As discussed previously, there is an opportunity to reduce energy consumption in Big Data processing. The state of the art and the related works ignored it. To show this opportunity, we aggregate 23 GB of data from four sources and divide it into 0.5GB blocks. We consider these blocks and show the average CPU utilization and processing time of them. Fig. 1 and Fig. 2 show average CPU utilization and processing time for various applications and different parts of input data. Experiments of the current section were run on an Intel Core-i7 4-core CPU at 2.8GHz with 4GB of RAM.

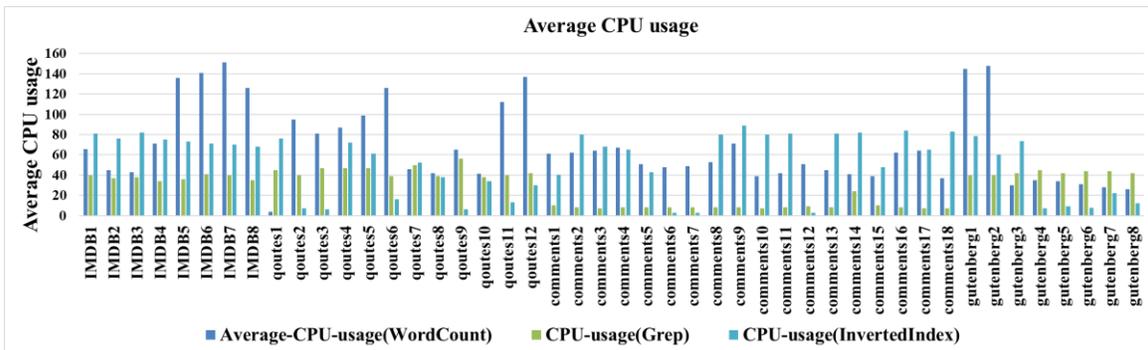

Fig. 1. CPU utilization in various parts of sources

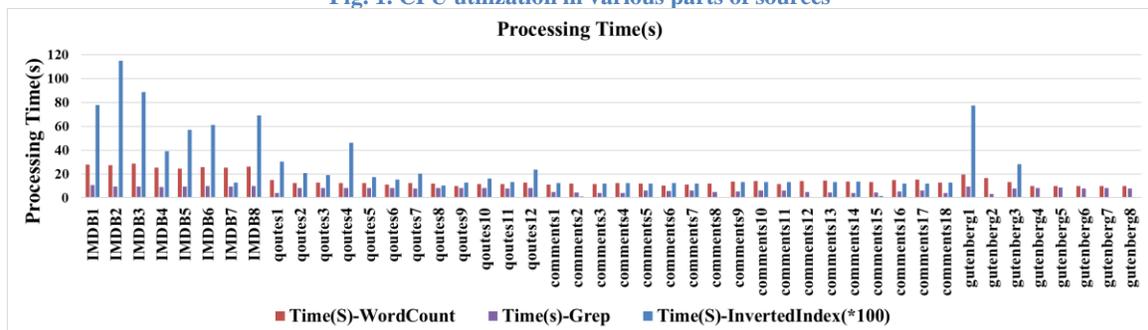

Fig. 2. Processing time for various parts of sources

CPU utilization and the processing time for IMDB, Quotes, Comments (Stack overflow), and Gutenberg have been presented in Fig. 1 and Fig. 2, respectively. Based on the results there is a wide variety in the processing requirements. For example, the data parts with a greater number of words require more CPU utilization and more processing time. This issue clearly shows that there is a novel potential for the reduction of energy consumption by using data variety.

We also have presented the results of motivational experiments in Table 1. Mentioned results, presents the average, variance, and coefficient of variation of CPU usages and processing time in each benchmark. Due to the variation values presented in Table 1, there is a significant opportunity to manage CPU utilization and power consumption.

Table 1. CPU utilization in three applications

| Applications | WordCount | Grep | Inverted Index |
|---|---|---|---|
| Average-CPU usage | 68% | 45% | 82% |
| Variance CPU usage | 42 | 17.5 | 30.5 |
| Coefficient of Variation of CPU usage | 0.51 | 0.65 | 0.64 |
| Average-average Processing time | 14.9 | 6.97 | 2283.5 |
| Variance of Processing Time | 31.3 | 4.12 | 6881847.7 |
| Coefficient of Variation of Processing Time | 2.1 | 0.6 | 3013.8 |

On the other hand, Due to the structure of MapReduce processing and 4Vs of big data, big data processing is a suitable area to apply the power reduction techniques such as DVFS.

**Approach.** Based on the [8], [9], and [2], we have shown that by the MapReduce processing, we can divide input data into some parts and manage it. We can process each data part with a different infrastructure or capacity. Furthermore, as we have shown in the previous work [8] data variety is one of the important features of big data, causes variation in resource consumption. This fact makes DVFS a suitable technique for the reduction of power/energy consumption in big data processing. To address the mentioned challenge; we present our power-conscious approach to managing the energy consumption of big data processing. As Fig. 3 shows, we use sampling to discover the input data. We have used pre-processing and an estimator to estimate the frequency and time of processing.

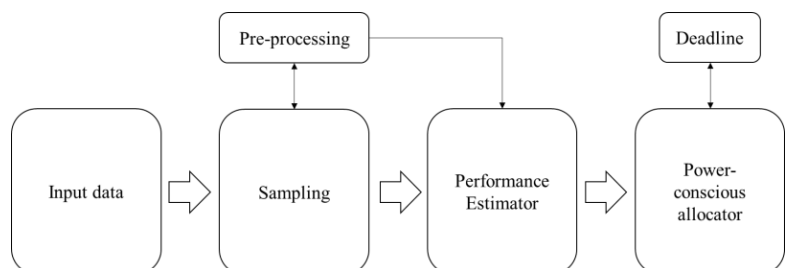

Fig. 3. our approach

**Contributions**. In this paper, we have the following contributions:

1. We have presented a framework to consider data variety for efficiently assigning resources in the big data processing.
2. We have used sampling to discover the amount of data variety.
3. We have implemented our approach in the Spark environment and evaluated it by some well-known datasets and applications.

Based on the presented contents, we should notice the following points in the current paper:
- ❖ Why using DVFS in big data processing?

1. Data variety causes a significant diversity in resource utilization. Variety oblivious approaches can lose the processing resources such as energy.
2. Using cloud computing for big data processing intensified the data variety and causes more variety in data. So, cloud providers and users must use techniques such as DVFS for reducing energy consumption.
3. Also data variety is one of the big data's 4V. Aggregating input data from various sources intensifies the data variety.
4. MapReduce is a well-known paradigm for big data processing. MapReduce consists of two main parts: Map and Reduce. Each phase of this paradigm has various impacts on the utilization of resources. By using this paradigm, we can divide input data into some data parts and process them with minimum overhead.

**Organization.** The rest of the paper is organized as follows: Section 2 presents an overview of the state of the arts and previous works. Section 3 describes the proposed approach and system design. The experimental result and evaluations are presented in section 4, and Finally, Section 5 includes the main conclusions and future works.

## 2. RELATED WORKS

Related works of our research are divided into 2 main categories. We have presented these categories in Fig. 4. The categories are:

1. Using Dynamic Voltage and Frequency Scaling (DVFS) to energy reduction.

2. Using other techniques to reduce energy consumption

**Using DVFS to energy reduction.** DVFS is a well-known approach to reduce energy reduction in case of a lack of energy. The authors in [6] used DFVS to reduce the energy consumption of MapReduce applications. They have compared their work with the default Spark Scheduler. We also have used the DVFS for reducing energy consumption in big data processing. Unlike our work, the mentioned research ignored the data variety. VM migration and scale down in case of low performance are considered in [10]. The authors in [11] have considered the variation of application requirements in big data in case of choosing cloud as an infrastructure for processing. Tuning CPU frequency based on the QoS has been presented in [12]. They have used a prediction method for adapting the frequency depends on the QoS and available time slot. They have reduced the energy consumption of the heterogeneous Hadoop cluster. The authors in [13] have used DVFS and machine learning approaches to reduce energy consumption in NoCs. The authors in [14] have used DVFS for microprocessors' power and energy reduction. DVFS based policies are used in [15] for the consolidation of virtual machines for energy-efficient cloud data centers. The authors in [16] have merged a thermal-aware approach and DVFS to manage the energy of the data center. In this paper, the factors of energy inefficiency are divided into two categories: resource underutilization and heat effects. The authors in [17] have used the DVFS technique for managing energy consumption of unknown applications. Using the DVFS technique for Fog-IoT applications has been considered in [18]. The authors achieve better QoS by using such techniques. The authors in [19] have used DVFS and approximation for reducing the processing cost. They have used the DVFS technique for each processing core to achieve better results and lower cost. Unlike our mentioned work, researchers have ignored the data variety. However, we have considered the data variety and have used the DVFS technique to achieve better results according to energy consumption. Some previous works in the area of energy and power consumption have been considered in [20] as the survey. The authors in [21] have considered DVFS as the main technique for presenting a QoS aware processing approach. They have reduced the energy consumption by presenting a two-stage approach to make a better decision about the processing frequency of each application. Edge computing and DVFS have been combined in [22] to achieve better results in terms of energy consumption. The authors have considered communication and computation and communication energy consumption. Like our work, the authors in [23] have divided the deadline into some time slots and assign suitable frequencies to the VMs in a way meeting the deadline and achieve better reliability.

**Using other techniques to reduce energy consumption.** The authors in [24] considered the server utilization to reduce energy consumption. They have also considered QoS in their problem. The authors in [25] have used a data-driven approach to improve the performance of HPC systems. The authors in [7] have presented a heuristic-based framework for energy reduction by load balancing. The authors in [3] and [4] have considered the energy consumption in MapReduce like distributed processing frameworks. The goal of these researches was to minimize energy consumption and guarantee the deadline by determining the number of worker nodes.

The impacts of failures and resource heterogeneity on the power consumption and performance of IaaS clouds have been investigated in [26]. The authors have used historical monitoring data from the online analysis of the host and network utilization without any pre-knowledge of workloads for the reduction of SLA-violation and energy reduction in [27].

The authors in [28] have considered application-level requirements for energy reduction. They have considered the effect of the variety of workloads on the utilization of VMs and network. They have reduced the energy cost by assigning a suitable amount of resources to the VMs. The authors in [29] survey the previous works on the energy consumption of data centers. They have divided the research areas into some parts and discussed them. They have not considered the data variety in their study. The authors in [30] have detected a variety in processing resources of the applications. They have selected suitable applications for processing in the edge. They have considered communication and computation overheads. We have also considered data variety and reduced the processing resources such as energy or cost [2], [9] , and [30].

This kind of research, like the first categories, have not considered the data variety. They also have not used the DVFS technique for reducing energy consumption.

3. Methods

In this section, we present the problem definition and the algorithm of the proposed approach. We have considered data variety in this method. Our problem is the reduction of energy consumption by applying DVFS to the computer nodes to overcome the inefficiency caused by data variety. For this reason, we divide the input data into some same size portions. We estimate the required processing resources for each portion by using sampling. Then, we select the suitable portions for applying DVFS techniques. In this problem, we must consider the deadline as a constraint. We have used the DVFS technique to reduce energy consumption and meet the deadline. For solving this problem, we have presented a heuristic approach. In this heuristic approach, we have used some notation for our presentation of the problem. Table 2 presents the notations that we used in this section.

**Table 2. Notations used in this paper**

| Notation | Description |
|---|---|
| D | Deadline |
| EC | Energy Consumption |
| FT | Finish time |
| UF | Utilize Factor |
| TS | Time Slot |
| $PT_i$ | The processing time of i-th block |
| TPT | Total Processing Time |
| $SFB_i$ | Suitable Frequency for processing $B_i$ |
| $U_i$ | Utilization of server i |
| $P_i$ | Processing power of server i |
| $N_{DP}$ | Number of Data Portions |
| $B_i$ | Data Block i |

**Problem Statement.** EC presents the energy consumption in this paper. We try to minimize the EC while the deadline should be met. So, the deadline is the constraint of our problem.

**Problem formulation.** The objective function to be minimized is the energy consumption and the constraint is the deadline.

$$Min(EC) \quad (1)$$

Subject to:

$$FT \leq D \quad (2)$$

(1) Presents the objective function and the

(2) Presents the constraint of our work.

To overcome the above problem, we have presented Algorithm 1. Before the presentation of the algorithm, we define a parameter "Utilize Factor".

$$P_i = (P_i^{full} - P_i^{idle}) * u_i^{CPU} + P_{idle} \quad (3)$$

$$u_i^{CPU} = UF_i * u_i^{full} \quad (4)$$

$$UF_i = PT_i / TS_i \quad (5)$$

$$\sum_0^N TS_i \leq Deadline \quad (6)$$

$$EC = \sum_0^N PT_i * P_i \quad (7)$$

Formula 3 to 6 calculate the Required Power for Processing (*RPC*) for each block. Formula 6 presents the constraint of the problem. Formula 7 calculates the energy consumption of processing.

**Our Algorithm.** Our algorithm is presented below.

Algorithm 1
---
1: **Input:** *Deadline,*
2: **output:** *SFB$_i$*
3:     **divide** *(Deadline into N$_{DP}$ Parts)*
4:     **divide** *(InputData into N$_{DP}$ Blocks)*
5: **while** *(TPT < D)*         // *Meeting the deadline*
6:     **while** *(Sample All Blocks)*   // *Until all blocks have been sampled*
7:         **Sample** *(B$_i$)*        // *To discover the amount of variety*
8:         **Estimate** *(SFB$_i$)*
9:     **end while**
10: **end while**

Lines 1-2 of Algorithm 1 is initializing the variables. Line 3 divides the deadline into some same size time slots. Line 4 divides the input data into some same size data block. As Fig. 5 shows each slot is assigned to one data block for processing. Based on the fixed size of time slots and data blocks, we can decide the frequency that is should be used to finish processing of the mentioned data portion in its time slot. Thus, only the data variety causes differences in frequencies. Line 7 uses sampling to discover the variety in initial blocks that are needed to estimate the required frequency for each block processing [2], [9]. Line 8 estimates the suitable frequency for processing of $B_i$ based on the time slot. The finishing time of processing should be lower than the deadline (line 5) and all data blocks must be sampled (line 6).

**Implementation.** In our approach, we divided the input data into some data blocks. In the Spark environment, these blocks are converted into some RDDs[1]. As Fig. 4 shows, we have used sampling to discover the amount of processing resources needed for processing each RDD. Based on this information we have decided the amount of resources needed for processing each RDD. As Fig. 5 shows, a certain frequency is assigned to each RDD. So, by applying this approach, we have used dynamic voltage and frequency scaling for big data processing.

---

[1] Resilient Distributed Datasets

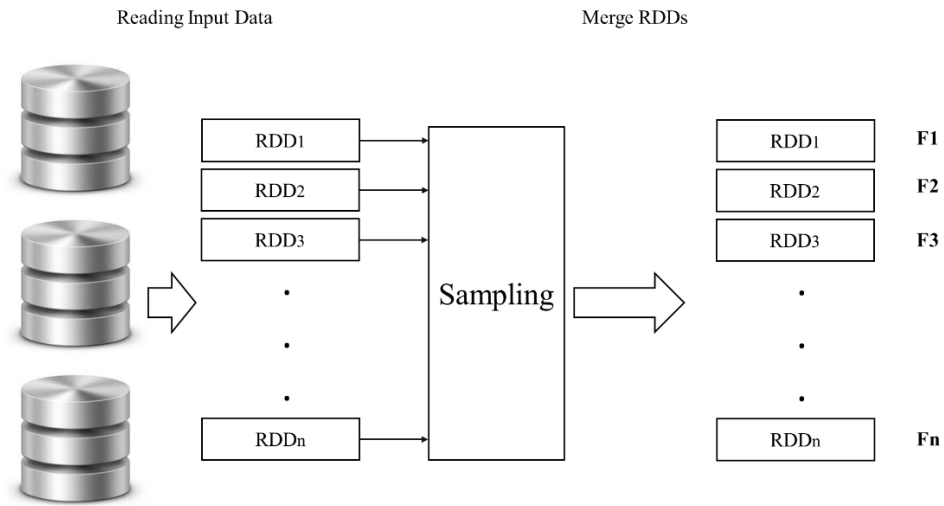

Fig. 4. Overview of our approach

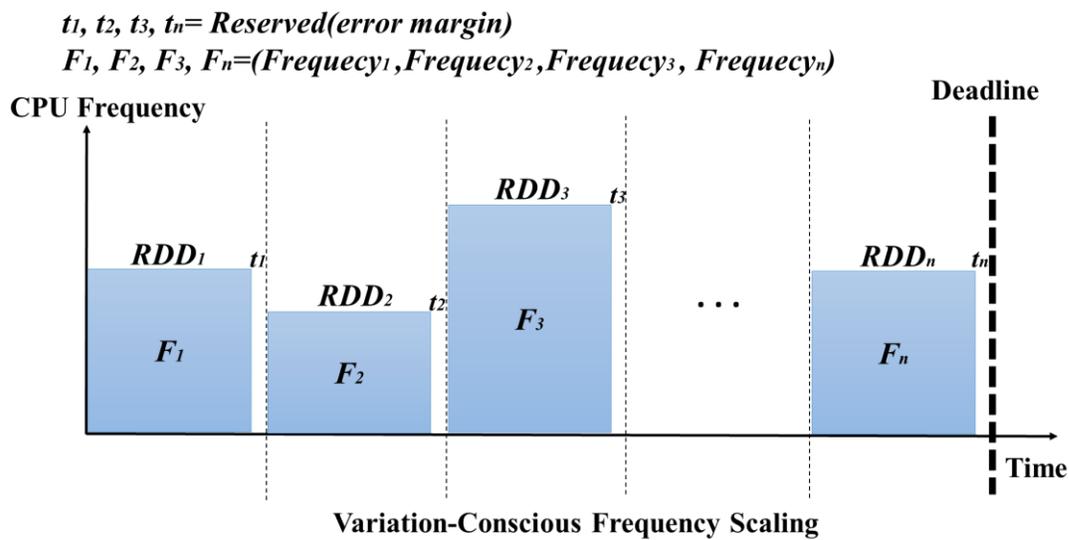

Fig. 5. Variation-Conscious Frequency Scaling (our approach)

As presented in Fig. 5 by using DVFS, we can apply various frequencies to the different RDDs based on their requirements to meet the deadline. For each time slot, we have considered a reserved area for error margin. These error margins can guarantee meeting the deadline.

## 4. RESULTS and DISCUSSION

We used three benchmarks from BigDataBench suite [32] in our evaluation process. We also have used TPC Benchmark (MAIL, SHIP, AIR, RAIL, TRUCK) and Amazon review dataset (Music, Books, Movies, Clothing, Phones) [33], [34]. Amazon product data contains product reviews and metadata from Amazon, including 142.8 million reviews spanning May 1996 - July 2014. TPC-H is a decision support benchmark. It consists of a suite of business-oriented ad-hoc queries and concurrent data modifications. The queries and the data populating the database have been chosen to have broad industry-wide relevance. This benchmark illustrates decision support systems that examine large volumes of data, execute queries with a high degree of complexity, and give answers to critical business questions. We have used four different sources [35], [36], [37] and Wikipedia for WordCount, Grep, Inverted Index, and AverageLength. We have used a bootstrapping method for generating 100GB data as input datasets [38]. Experiments were run on three machines, Intel Core-i7 4-core CPU at 2.8 GHz with 4 GB of RAM. We apply the DVFS to some parts of data and reduce CPU frequency to the 1.6 GHz.

**Applications.** Applications are as follows:

- **WordCount**: This application Counts the number of words in the file.

- **Grep**: It searches and counts a pattern in a file.

- **Inverted Index**: This application is an index data structure storing a mapping from content to its locations in a database file.

- We also consider **AVG** (average) for TPC-H datasets and **SUM** for Amazon datasets.

**Comparison**. We have compared our approach with a default scheduler of Spark [6]. In this approach, the same amount of resources is given to each application. In this kind of frequency scaling, we have considered a fixed frequency as CPU frequency (i.e., default Spark scheduler). This approach ignored the data variety, so we named it **Data Variety Oblivious (DVO)**.

Fig. 6 to Fig. 10 depicts the execution time and energy consumption of approaches. Processing time and processing energy consumption of Wordcount, Grep, Inverted Index, AVG (TPC benchmark), and SUM (Amazon benchmark) have been presented in Fig. 6, Fig. 7, Fig. 8, Fig. 9 and Fig. 10, respectively. Processing time and energy consumption have been shown in red and purple, respectively.

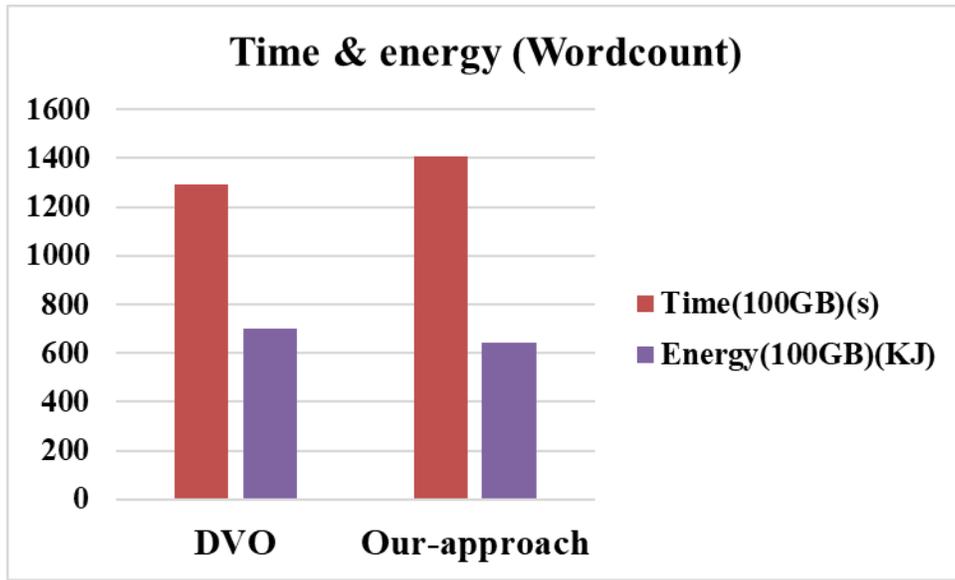

Fig. 6. Processing time and energy of Wordcount

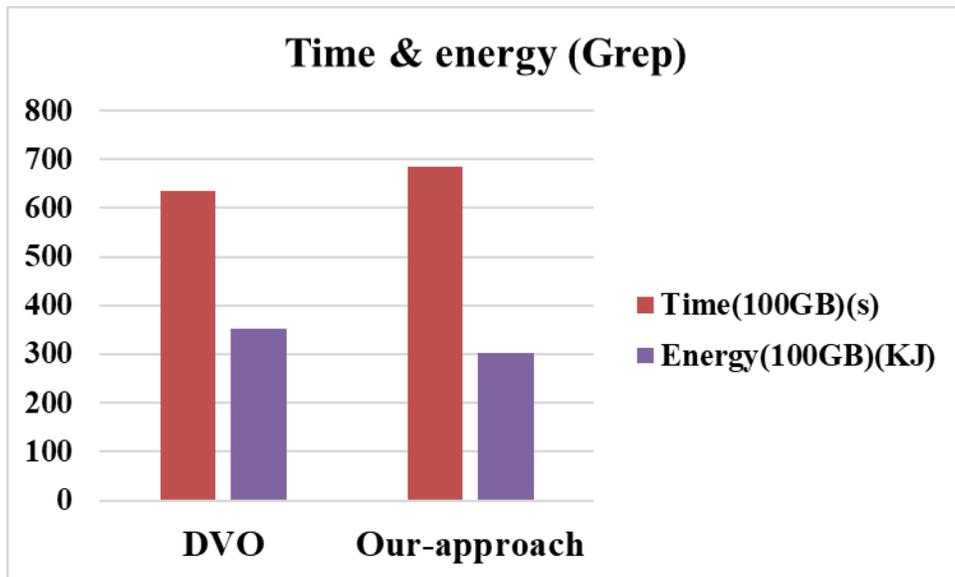

Fig. 7. Processing time and energy of Grep

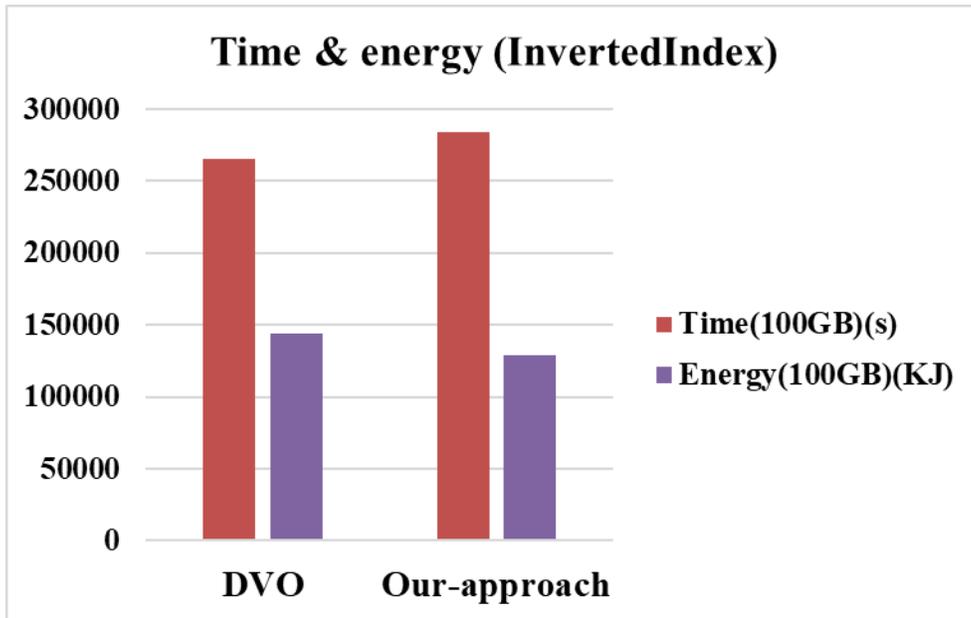
Fig. 8. Processing time and energy of Inverted Index

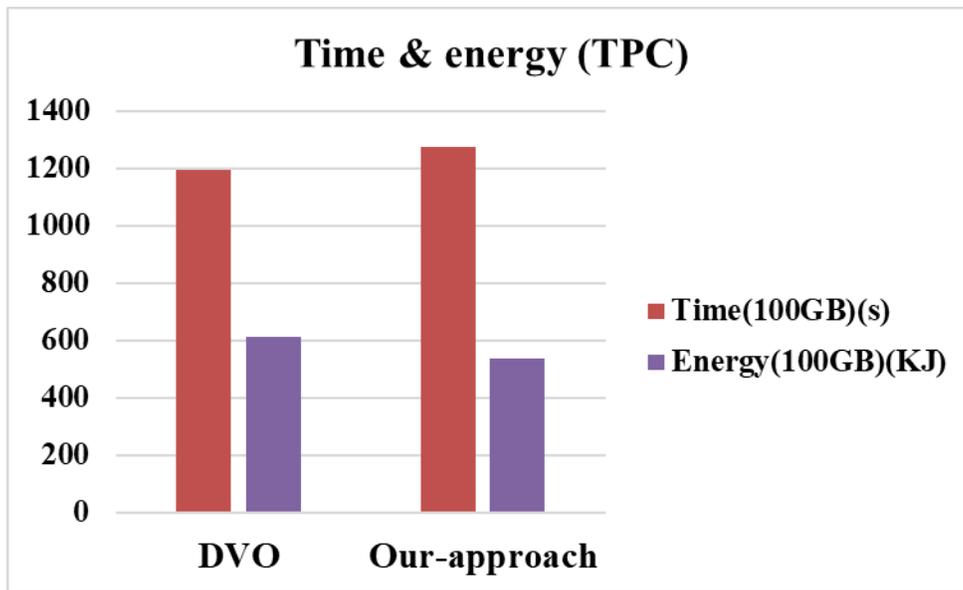
Fig. 9. Processing time and energy of TPC Datasets

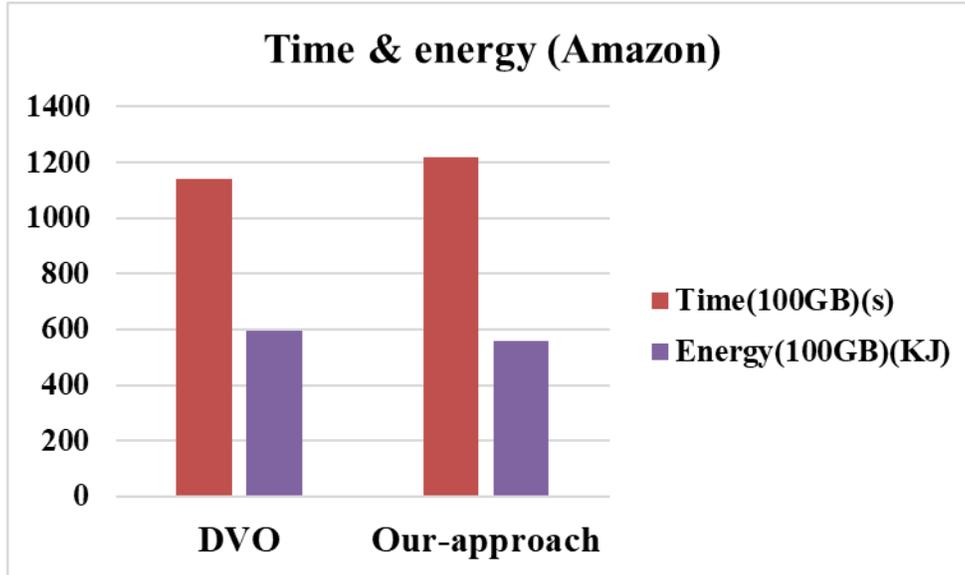

**Fig. 10. Processing time and energy of Amazon datasets**

As can be seen, our proposed approach can surpass the other in all the applications in terms of energy consumption. Based on the deadline as a constraint, we have delayed the completion of processing. We have met the deadline in all applications. Based on the results presented in Fig. 6 to Fig. 10, our approach can surpass the default scheduler and achieve 9%, 15%, 11%, 13%, and 7% improvement for energy consumption in Wordcount, Grep, Inverted Index, TPC, and Amazon benchmarks. As Fig. 6 to Fig. 10 show our approach increases the processing time by 8%, 7%, 6%, 7%, and 6% in Wordcount, Grep, Inverted Index, TPC, and Amazon, respectively. This increases in the processing time should be considered for meeting the deadline. We have analyzed this issue in the sensitivity analysis in the following sections.

**Sensitivity analysis.**

We also analyze the impact of data variety and the deadline for the performance of our work. For modeling data variety, we have used a mathematic law and for the deadline, we have considered two conditions.

**Sensitivity to the data variety.** Aggregating data from multiple sources causes uneven distribution. Uneven distribution intensifies data variety among data. In case of increasing data variety, our approach is able to save more energy and we have a better choice to apply the DVFS technique. We have considered two types of data variety for our experiments: Moderate and High.

**Modeling data variety.** We have used Zipfian [39], [40] distribution to generate a variety of data. Zipf's law states that out of a population on N elements, the frequency of elements of rank k, $f(k;z,N)$ is:

$$f(k; z, N) = \frac{\frac{1}{k^z}}{\sum_{n=1}^{N}(\frac{1}{n^z})}$$

Following the Zipfian distribution, the frequency of occurrence of an element is inversely proportional to its rank.
In the current context, let:
1) N = total number of input partitions;
2) k be their rank; partitions are ranked as per the number of records in the partition that satisfy the given predicate;
3) z be the value of the exponent characterizing the distribution.
We have considered z=0 for uniform distribution and z=2 for high variety.

Fig. 11 and Fig. 12 present the impact of data variety on processing time and energy consumption. The horizontal axis shows the benchmarks and the vertical shows the normalized processing time and energy consumption. The processing time and energy consumption are normalized to the Data Variety Oblivious approach. Moderate data variety (z=1 in Zipfian distribution) have been considered in Fig. 11 and high data variety (z=1 in Zipfian distribution) have been considered in Fig. 12. As shown in Fig. 11 and Fig. 12, when data variety increased, our approach can perform better results in terms of energy consumption.

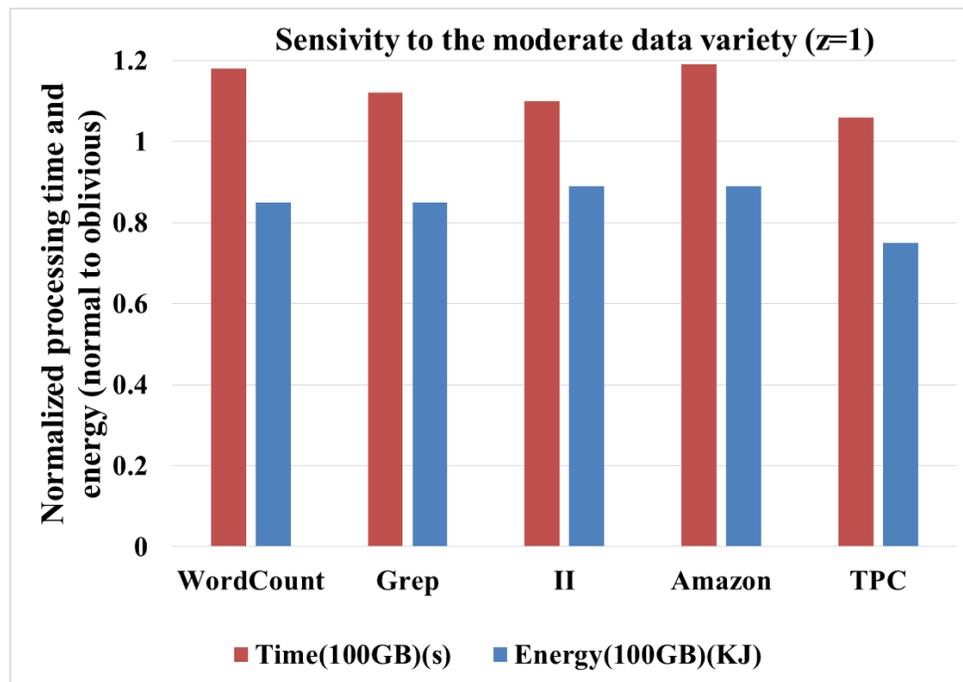

Fig. 11. Sensitivity analysis to data variety (z=1)

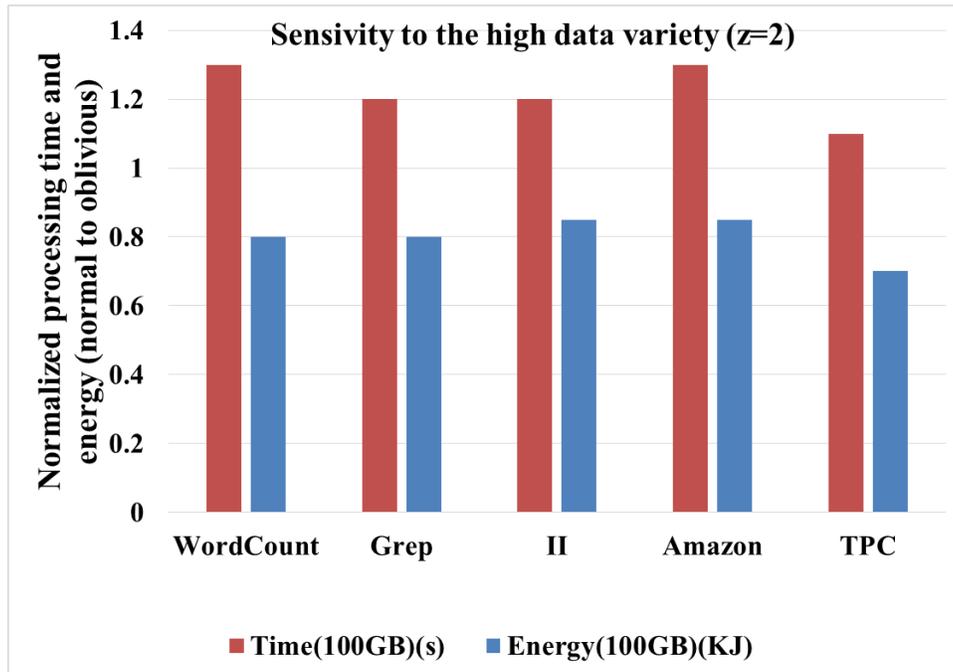
Fig. 12. Sensitivity analysis to data variety (z=2)

**Sensitivity to the Deadline.**

We have considered two statuses for the deadline, the tight deadline, and the firm deadline. We have presented these statuses in Table 3. A tight deadline is less than the firm deadline. In case of a tight deadline, the processing should be done at a higher speed.

While there is a tight deadline, we have limited choice to apply DVFS to the computer node. Our approach has better performance in case of the firm deadline. In other words, in case of firm deadline, we can apply the DVFS technique to more parts of data in comparison to the tight deadline. So, our approach can generate better results. We have shown the two conditions of the deadline in Table 3.

Table 3. Tight and Firm Deadline for benchmarks

| Benchmarks | Tight Deadline(s) | Firm Deadline(s) |
| --- | --- | --- |
| Wordcount | 1350 | 1500 |
| Grep | 670 | 730 |
| Inverted Index | 27000 | 30000 |
| TPC | 1250 | 1400 |
| Amazon | 1150 | 1350 |

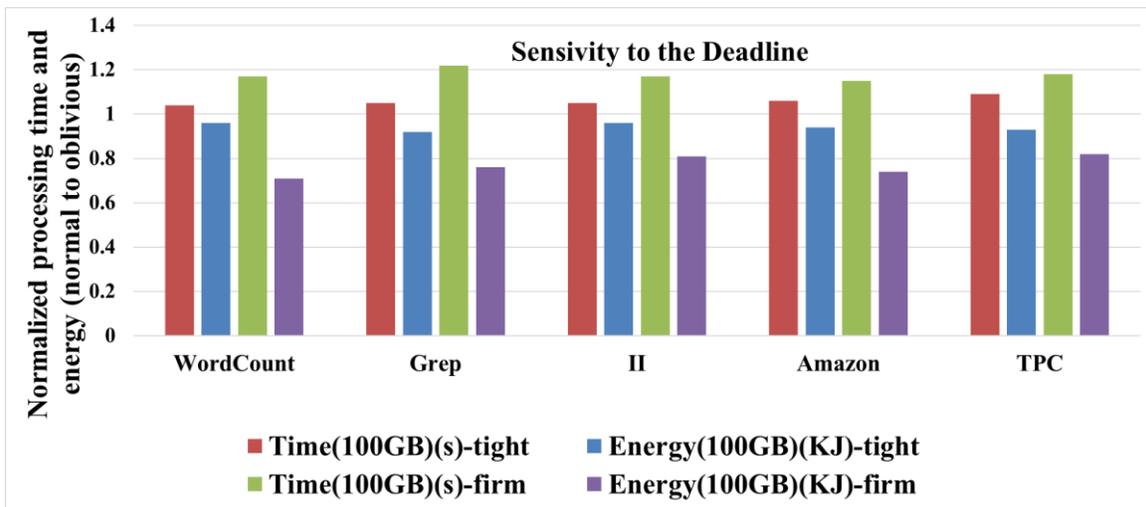
**Fig. 13. Sensitivity analysis to Deadline**

As Fig. 13 shows, our approach has better performance in firm deadline. In the firm condition, we have an opportunity to use the DVFS technique more than the tight deadline. This issue causes more improvement in energy consumption in comparison to the tight condition.

At the end of this section, we should discuss two important issues: Overhead and the Usages

- ❖ **Discussion on the overhead.** Our approach is a very low overhead solution. Sampling has less than 1% overhead for generating a 5% error margin and a 95% confidence interval. For this issue, we have a wide approach and description in [9].

- ❖ **Discussion on the usages.**
  - This approach is applicable for cloud service provider and every cloud user that can manage the infrastructure.
  - Based on the variety that is one of the features of big data, this approach could be used for processing big data applications.
  - This approach reduces energy consumption and the cost of energy. So, cloud providers clearly can benefit from it.
  - In this paper, we have presented an approach for reducing energy consumption in Big data processing for accumulative applications. We have presented the definition of accumulative application in [9]. This type of applications is an important type of Big Data applications [8], [9].

## 5 CONCLUSION

In summary, we have studied the impact of data variety on energy consumption via controlling CPU utilization in the Big Data processing. In the first step, we divide input data into some same size blocks.

Then, we have used sampling to estimate the processing resource needed for each block. Finally, we have processed the data blocks with the DVFS technique. The results show that our variety- conscious approach produced better results in comparison to data variety oblivious approach. Based on the results, in firm deadline, our approach generates better results compared with tight conditions. Because, we are able to apply the DVFS technique to more parts of the data in the mentioned condition.

Many interesting directions exist to continue from this work. First, considering energy cost in various parts of data and geographical area. Based on this idea, we can process input data when/ where the energy cost is minimum and improve the big data processing cost. Second, we can consider renewable energy for reducing energy consumption. So, we can process the main part of the input data by more efficient and lower cost energies.

Abbreviations

**D:**     Deadline
**EC**:    Energy Consumption
**FT**:    Finish time
**UF**:    Finish time
**TS**:    Time Slot
**$B_i$**:    The i-th block
**$PT_i$**:   The processing time of i-th block
**RPC**:   Required Power for Processing
**REP**:   Required Energy for Processing
**$SFB_i$**: Suitable Frequency for processing $B_i$
**AVG**:   Average
**$U_i$**:    Utilization of server i
**$P_i$**:    Processing power of server i
**DVO**:   Data Variety Obvious

**Declarations**

**Authors' contributions**

HA is the primary researcher for this study. His contributions include the original idea, literature review, implementation, and initial drafting of the article. FF discussed the results with the primary author to aid writing of the evaluation and conclusion sections and played an essential role in editing the paper. MF help to improve the research concept and played a crucial role in the research. All authors read and approved the final manuscript.

**Acknowledgements**

Not applicable.

**Competing interests**

The authors declare that they have no competing interests.

**Availability of data and materials**

BigDataBench: http://prof.ict.ac.cn/.

TPC Benchmark: http://www.tpc.org/information/benchmarks.asp.

Amazon product data: http://jmcauley.ucsd.edu/data/amazon/.

IMDB data files: https://datasets.imdbws.com/.

Gutenberg datasets: https://www.gutenberg.org/.

Quotes-dataset: https://www.kaggle.com/akmittal/quotes-dataset.

**Funding**

Not applicable.

**Ethics approval and consent to participate**

Not applicable.

**Consent for publication**

Not applicable.